\documentclass[%
 reprint,
 amsmath,amssymb,
 aps,
floatfix,
]{revtex4-2}
\usepackage{graphicx}
\usepackage{dcolumn}
\usepackage{bm}
\usepackage{amsmath}

 \usepackage{verbatim}

\usepackage[colorlinks,linkcolor=blue,citecolor=blue,urlcolor=blue,hyperindex,breaklinks]{hyperref}
\allowdisplaybreaks[4]
\usepackage{subfigure}
\allowdisplaybreaks[4]
\usepackage{subfigure}
\usepackage{xcolor}
\usepackage{soul}
\soulregister\cite7
\soulregister\ref7

\begin{document}
\title{Quantum entanglement and Einstein-Podolsky-Rosen steering in ultrastrongly light-matter coupled system
}
\author{Yu-qiang Liu$^{1, 2}$}
\email{Electronic address: liuyuqiang@htu.edu.cn}
\author{Shan, Sun$^1$}
\author{Yi-jia, Yang$^2$}
\author{Zheng, Liu$^2$}
\author{Xingdong, Zhao$^1$}
\author{Zunlue, Zhu$^1$}
\author{Wuming, Liu$^3$}
\author{Chang-shui Yu$^{2}$}
\email{Electronic address: ycs@dlut.edu.cn}
\affiliation{$^1$School of Physics, Henan Normal University, Xinxiang 453007, China}
\affiliation{$^2$School of Physics, Dalian University of Technology, Dalian 116024,
P.R. China}
\affiliation{$^3$Beijing National Laboratory for Condensed Matter Physics, Institute of Physics, Chinese Academy of Sciences, Beijing 100190, China}

\date{\today}

\begin{abstract}

This work presents a scheme for engineering quantum entanglement and Einstein-Podolsky-Rosen (EPR) steering with
Gaussian measurements based on the quantum Hopfield model that incorporates a common thermal reservoir. We begin by examining quantum correlations, specifically quantum entanglement and EPR steering, in the ground state. These quantum correlations primarily stem from squeezing interactions in weak and normal strong coupling regimes. As the coupling strength increases, especially upon entering the ultrastrong coupling regime, the correlations emerge from the combined effect of squeezing and mix-mode interactions. Importantly, this scenario enables the realization of two-way EPR steering. Moreover, lower optical frequencies enhance both quantum entanglement and EPR steering. Further, when considering thermal effects, the ultrastrong and deep strong coupling regimes, paired with lower optical frequencies, lead to improved entanglement. The one-way EPR steering for resonant case can be effectively controlled in the ultrastrong and deep strong coupling regimes which originates from the asymmetry of subsystem and reservoir coupling induced by the diamagnetic term. Additionally, one-way EPR steering can also be produced for nonresonant case. In this case, the asymmetry of the subsystem and reservoir originates from the combined effect of nonresonant frequencies and diamagnetic term. Our findings have the potential to inspire further research into quantum information processing that leverages light-matter entanglement and EPR steering.

\end{abstract}
\maketitle
\section{Introduction}

Recent advancements in cavity quantum electrodynamics (QED) with ultrastrong coupling \cite{RevModPhys.91.025005, frisk2019ultrastrong} have positioned it as a vital platform for exploring a range of intriguing phenomena. The ultrastrong coupling regime shows that the coupling $g$ becomes a considerable fraction of the bare light (matter) frequency $\omega_a$ ($\omega_b$) of uncoupled systems, typically $0.1 \leq g/\omega_{a, b}< 1$. Moreover, as the coupling is further increased, the deep strong coupling regime can be obtained when $g/\omega_{a, b}> 1$ \cite{RevModPhys.91.025005, frisk2019ultrastrong}. Among these are dynamical Casimir-like effects \cite{PhysRevLett.98.103602}, light–matter decoupling \cite{PhysRevLett.112.016401}, nontrivial ground state \cite{PhysRevB.72.115303} and quantum Rabi physics \cite{koch2023quantum}. Furthermore, advancements in quantum simulation techniques \cite{PhysRevA.87.033814} provide a pathway to explore the physics associated with the ultrastrong coupling regime. The realization of the ultrastrong coupling regime of light and matter has been demonstrated in various systems, such as plasmonic nanoparticle crystals \cite{mueller2020deep}, cavity QED setups \cite{PhysRevLett.124.040404, PhysRevA.98.053819} and circuit QED architectures \cite{PhysRevB.95.224515, RevModPhys.93.025005}. As the coupling strength between light and matter increases, new hybrid modes, referred to as "polaritons," emerge \cite{flick2018strong}. These include magnon–polaritons \cite{PhysRevB.97.014419}, exciton–polaritons \cite{khitrova2006vacuum}, and Landau polaritons \cite{bayer2017terahertz}. 

The Hopfield model \cite{PhysRev.112.1555} serves as an effective framework for describing the properties of these hybridized polaritons while investigating various forms of light-matter coupling, such as magnon-photon interactions, coupled molecular vibrations with microcavity modes, and plasmon-microcavity coupling. The Hopfield model effectively describes the properties of hybridized polaritons and allows for exploring various light-matter coupling. This includes magnon-photon coupling \cite{PhysRevApplied.20.024039}, coupled molecular vibrations and microcavity modes \cite{barra2021microcavity}, plasmon-microcavity coupling \cite{baranov2020ultrastrong}, interactions between Landau-quantized 2D electrons and terahertz cavity photons \cite{zhang2016collective}, as well as the coupling of photons and plasmons \cite{mueller2020deep}. Additionally, the model encompasses the interaction of coupled vibrational modes of molecules \cite{Pino_2015}, coupled microwave resonators \cite{PhysRevA.103.023707}, coupled nanomechanical oscillators \cite{RevModPhys.86.1391}, and magnetomechanical systems \cite{PhysRevB.97.024109}. The Hopfield-type model has been extensively studied concerning phenomena such as the dissipation and detection of polaritons \cite{PhysRevA.86.063831}, the breakdown of the Purcell effect \cite{PhysRevLett.112.016401}, quantum phase transitions \cite{nataf2010no, PhysRevLett.107.113602}, and the quantum estimation of the diamagnetic term \cite{rossi2017probing}.
\begin{figure}
\centering
\subfigure[]{
\includegraphics[width=0.85\columnwidth]{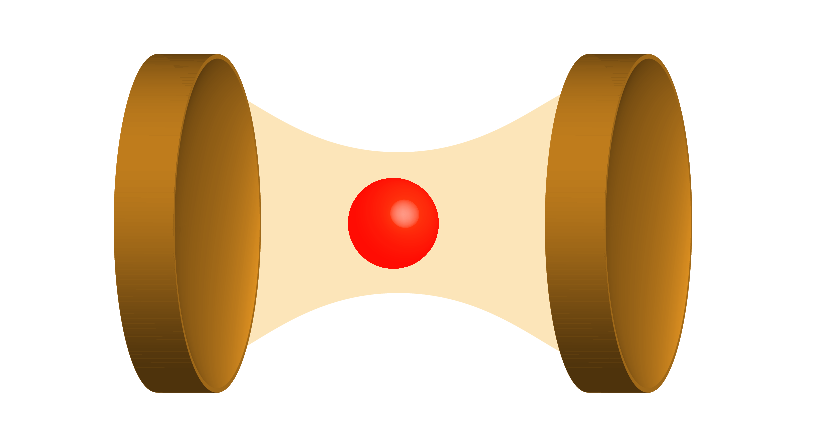}}
\centering
\subfigure[]{
\includegraphics[width=0.85\columnwidth]{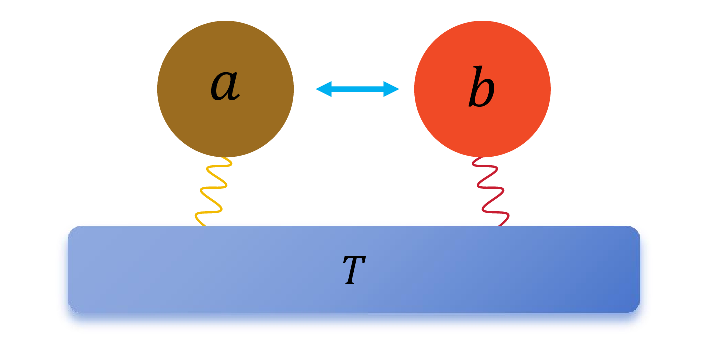}}
\caption{\label{fig: model}(a) A diagram of two ultrastrongly coupled oscillators consisting of a cavity photon mode and a matter mode; (b) The two coupled oscillators with modes $a$ and $b$, sharing a common thermal reservoir at temperature $T$.}
\end{figure}

In recent years, there has been a growing interest in advancements related to quantum correlations, particularly in their theoretical exploration and practical applications for quantum information processing. Quantum entanglement, one of the most thoroughly studied forms of quantum correlation, can execute various quantum tasks \cite{li2013entanglement, yin2020entanglement, malia2022distributed}. Over the past few decades, quantum entanglement has also been extensively researched in the field of quantum cavity optomechanics \cite{RevModPhys.86.1391}. By drawing parallels to cavity optomechanical systems, the cavity magnomechanical system \cite{zhang2016cavity} — which features yttrium iron garnet (YIG) material known for its remarkable versatility — has emerged as a promising platform for various applications, including quantum memories \cite{PhysRevLett.107.133601} and tunable microwave filters and amplifiers \cite{bergeal2010phase}. The generation of quantum entanglement can be manipulated by coupling magnons with other quasiparticles \cite{PhysRevLett.127.087203, PhysRevLett.124.053602}. EPR (Einstein-Podolsky-Rosen) steering, a strict subset of quantum entanglement that is stronger than entanglement but weaker than Bell nonlocality, has numerous potential applications, including quantum-key distribution(QKD) \cite{PhysRevA.85.010301, Walk16}, subchannel discrimination \cite{PhysRevLett.114.060404} and secure quantum teleportation \cite{PhysRevLett.115.180502}. The Gaussian one-way steering has first been described by introducing additional amounts of loss to one mode of a two-mode squeezed state \cite{WOS:000308695100011}. The asymmetry steering can be achieved by obtaining the asymmetry states, which is realized by adding different losses or noises to the subsystems at the cost of reducing steerability \cite{WOS:000308695100011, WOS:000349934700022, PhysRevA.95.052114}. Other methods also have proposed to control the asymmetry, such as controllability of the
amplitude of signals injected into a nondegenerate parametric
oscillator \cite{PhysRevLett.119.160501}, manipulation of the squeezing degree or the environmental temperature \cite{PhysRevA.98.042115}, and modulation of the ratio of coherent information exchange frequencies between two Kittel modes and the cavity mode \cite{PhysRevApplied.15.024042}. So far, quantum steering has been investigated in various quantum systems, such as cavity optomechanical systems \cite{PhysRevA.98.042115, PhysRevA.88.052121, PhysRevA.107.013507, PhysRevA.109.023532, PhysRevA.95.053842}, cavity-magnonics systems \cite{PhysRevLett.127.087203, PhysRevApplied.15.024042}, and cavity-magnomechanical systems \cite{PhysRevLett.121.203601, PhysRevA.103.053712}. Although significant researches have been conducted on quantum entanglement and EPR steering, most proposals primarily focus on the weak and normal strong coupling regimes of light and matter, with limited exploration of the ultrastrong and deep strong coupling regimes. Furthermore, the effects of ultrastrong and deep coupling regimes on generating or enhancing quantum correlations remain unclear. This work will examine quantum entanglement and EPR steering in the context of two coupled oscillators operating within the ultrastrong and deep strong coupling regimes.

The paper is organized as follows in Section \ref{Sec. II}, we introduce the quantum Hopfield model for light-matter interaction. To describe the system's dynamics, we derive the master equation based on the standard method and introduce the measures of quantum entanglement and EPR steering. In Sec. \ref{Sec. III}, we provide the main results and critical observations. We first briefly characterize the quantum entanglement and EPR steering of two ultrastrongly coupled bosonic modes in the ground state. We also discuss the impact of thermal effects on quantum correlations. The conclusion is given in Sec. \ref{conclusion}.


\section{Physical model and dynamics} \label{Sec. II}

We will consider a model that consists of coupled photonic mode $a$ and bosonic matter mode $b$, as shown in Fig. \ref{fig: model}. The interaction Hamiltonian can be viewed as a Hopfield-type model and is given (in units of $\hbar=1$ and $k_{B}=1$) \cite{PhysRevApplied.20.024039, PhysRevLett.112.016401, PhysRevA.91.063840, baranov2020ultrastrong, PhysRevApplied.20.024039, mornhinweg2024mode, halbhuber2020non}
\begin{align} \label{H_{S}}
H_{S} =\omega_{a} a^{\dagger} a+\omega_{b} b^{\dagger} b+\lambda (a+a^{\dagger})(b+b^{\dagger})+D (a^{\dagger}+a)^2,
\end{align}
where $\lambda$ and $D=\frac{\lambda^2}{\omega_{b}}$ denote the interaction strength of light-matter and the so-called diamagnetic term. 
As shown in Ref. \cite{PhysRevA.103.023707}, the Hamiltonian (\ref{H_{S}}) can be extended in a general form,
\begin{align} \label{H-S}
\begin{split}
H_{S} &=\omega_{a} a^{\dagger} a+\omega_{b} b^{\dagger} b+\lambda_{1}(a^{\dagger} b+a b^{\dagger})+\lambda_{2} (a^{\dagger} b^{\dagger}+a b) +\\&+D (a^{\dagger}+a)^2,
\end{split}
\end{align}
where $\lambda_1$ and $\lambda_2$ denote the mode-mixing and mode-squeezing interactions, respectively. In order to understand the mechanism of EPR entanglement, one can rewrite Eq. (\ref{H-S}) as the following form,
\begin{align} \label{H_{SS}}
\begin{split}
H_{S} &=(\omega_{a}+2 D) a^{\dagger} a+\omega_{b} b^{\dagger} b+\lambda_{1} (a^{\dagger} b + a b^{\dagger})+ \\&\lambda_{2} (a^{\dagger} b^{\dagger} + a b)+ D (a^{\dagger}  a^{\dagger}+a a),
\end{split}
\end{align}
where the terms in $\lambda_1$, $\lambda_2$, and $D$ describe the 
beam-splitter, two-mode squeezing-type interaction, and single-mode squeezing operators, respectively. The two-mode squeezing-type interaction can generate the EPR entanglement \cite{PhysRevA.40.913}.  
Additionally, it is noted that $\lambda_{1}= \lambda_{2}$ for the interaction between light and natural atoms, while the case of $\lambda_{1} \neq \lambda_{2}$ has also been explored in Refs. \cite{makihara2021ultrastrong, PhysRevX.4.021046}. Given that the bosonic operators are bilinear, the Hamiltonian (\ref{H_{S}}) can be diagonalized by employing the Hopfield transformation \cite{PhysRev.112.1555},
\begin{equation}
H_{\mathrm{Hop}}=\sum_{j=U, L} \omega_{j} p_j^{\dagger} p_j,
\end{equation}
where $\omega_{j}$ represents the polariton frequencies. $p_{j}$ is a linear combination of the light and matter operators, namely, $p_{j}=w_{j} a+x_{j} b+y_{j} a^{\dagger}+z_{j} b^{\dagger}$ and the related coefficients $w_{j}$, $x_{j}$, $y_{j}$ and $z_{j}$ are given in Appendix \ref{Appendix A}. Moreover, $p_{j}$ satisfies the relation $[p_{j}, p^{\dagger}_{i}]=\delta_{j, i}$, where $\delta_{j, i}$ is the Kronecker symbol. 
Similarly, one can utilize the polariton operator $p_{j}$ to represent the original operators $a$ and $b$,
\begin{align}  \label{original operator}
\begin{aligned}
&a=w_{U}^* p_{U}+w_{L}^* p_{L}-y_{U} p_{U}^{\dagger}-y_{L} p_{L}^{\dagger}, \\
&b=x_{U}^* p_{U}+x_{L}^* p_{L}-z_{U} p_{U}^{\dagger}-z_{L} \hat{p}_{L}^{\dagger}.
\end{aligned}
\end{align}
 The polariton frequencies can be expressed as 
$\omega_{U/L}=\sqrt{\frac{\omega_a^2+4 D \omega_a+\omega_b^2}{2} \pm \sqrt{\left(\frac{\omega_a^2+4 D \omega_a-\omega_b^2}{2}\right)^2+4 \lambda^2 \omega_a \omega_b}}$. Noted that $\omega_U \omega_L=\omega_a \omega_b$ with $D=\frac{\lambda^2}{\omega_b}$. When $D=0$, this system can reduce to the form of two bosonic coupled oscillators \cite{PhysRevA.86.012316}, with the polariton frequencies $\omega^{\prime}_{U/L}=\sqrt{\frac{\omega_a^2+\omega_b^2}{2} \pm \sqrt{\left(\frac{\omega_a^2-\omega_b^2}{2}\right)^2+4 \lambda^2 \omega_a \omega_b}}$. There exists a critical point $\lambda_C=\frac{\sqrt{\omega_{a} \omega_{b}}}{2}$ that is dependent on the Hopfield-Bogoliubov matrix \cite{nataf2010no}. Crucially, the low excitation energy $\omega_{L}$ is real only $\frac{\omega_a^2+\omega_b^2}{2} \geq \sqrt{\left(\frac{\omega_a^2-\omega_b^2}{2}\right)^2+4 \lambda^2 \omega_a \omega_b}$ or equivalently $\lambda<\frac{\sqrt{\omega_{a} \omega_{b}}}{2}=\lambda_{C}$ similar to Refs. \cite{PhysRevResearch.4.013171, PhysRevA.86.012316}. When the coupling strength $\lambda>\lambda_{C}$, the polariton mode becomes dynamically unstable, causing $\omega_{L}$ to become complex-valued. Additionally, the relevance of various quantum master equations in describing dissipative critical behavior has been questioned \cite{PhysRevResearch.4.013171}.

\subsection{Coupling to the thermal environment}
To effectively characterize the loss in an open system under ultrastrong and deep strong coupling of light and matter, two key variables must be introduced: the characteristics of the environment and the nature of the interaction between the system and the environment. The environment is represented as ensembles of harmonic oscillators, as outlined in Ref.\cite{RevModPhys.59.1},
\begin{equation}
H_{R}=\sum_{k} \omega_{k}{c}_{k}^{\dagger}{c}_{k},
\end{equation} 
where $c_{k}$, $c^{\dagger}_{k}$, and $\omega_{k}$ refer to the annihilation and creation operators and frequency of the reservoir modes, respectively. The annihilation and creation operators satisfy the commutation relation $[c_{k}, c^{\dagger}_{l}]=\delta_{k, l}$. When considering light-matter interactions in the regime of ultrastrong coupling, it is important for the system-reservoir coupling to retain the counter-rotating terms \cite{PhysRevA.89.023817}. We designate the two oscillators with modes $a$ and $b$ as subsystems $A$ and $B$, respectively. The interaction Hamiltonian of the system with a common reservoir can be expressed as follows
\begin{equation}
H^{C}_{I} = \sum_k [ f_{A k} (a^{\dagger}+a)+ f_{B k} ( b^{\dagger}+b) ](c^{\dagger}_{k} +c_{k}),
\end{equation}
where $f_{\nu k}$ with $\nu=A, B$ represents the coupling strengths between the $\nu$-th system and the reservoir modes.
Utilizing Eq. (\ref{original operator}), this interaction Hamiltonian can be rewritten as
\begin{align} \label{Eq.7}
H^{C}_{I} =(p^{\dagger}_{U} C_{U}+ p_{U} C^{\dagger}_{U})+ (p^{\dagger}_{L} C_{L}+ p_{L} C^{\dagger}_{L}), 
\end{align}
together with $C_{j}=\sum_k [f_{A j} W_j +f_{B j} X_j ](c^{\dagger}_{k}+c_{k})$, $W_j=w_j-y_j$ and $X_j=x_j-z_j$.

\subsection{Quantum master equation} 

As quantum systems inevitably lose energy to the external environment, the evolution is no longer unitary. The dynamical evolution of an open quantum system can be characterized in the interaction picture by a global master equation based on the Born-Markov-secular approximation \cite{breuer2002theory}. When considering the system coupled to a common reservoir at temperature $T$, we can derive the corresponding master equation in the interaction picture as follows,
 \begin{align}
 \begin{split}
 \label{eq:common_ms}
\frac{d \rho}{dt}&=\mathcal{L}^{C }_{A} [\rho]+\mathcal{L}^{C}_{B} [\rho]+\mathcal{L}^{C}_{A B} [\rho] \\ 
&=\sum_{j=U, L}\lbrace \Gamma^{C }(- \omega_{j}) \mathcal{D}[p_{j}] \rho +
\Gamma^{C}(\omega_{j}) \mathcal{D}[p^{\dagger}_{j}] \rho \rbrace,
\end{split}
\end{align}
where $\Gamma^{C}(\pm \omega_{j})=|\sqrt{\Gamma^{C}_{A}(\pm \omega_{j})} W_{j}+\sqrt{\Gamma^{C}_{B}(\pm\omega_{j})} X_{j}|^2$ for simplification. The dissipator is the standard Lindblad operator, which is expressed as $\mathcal{D}[o] \rho=o \rho o^{\dagger}-\frac{1}{2}\left\{o^{\dagger} o, \rho\right\}$. The spectral densities are given by $\Gamma_{\nu} (\omega)=\eta^{\nu}(\omega) N(\omega)$, and the thermal occupation number is defined as $N(\omega)=\frac{1}{e^{ \omega/ T}-1}$. Here, we consider the Ohmic reservoir spectral-response functions,
$\eta^{\nu}(\omega) =\gamma^{\nu} \omega$ with $\omega\geq 0$.

\subsection{Quantum entanglement and EPR steering}

The analytic formula can characterize Gaussian states of two bosonic modes. Therefore, we can define hermitian quadrature operators as $d_{j}=\frac{p_{j}+p_{j}^{\dagger}}{\sqrt{2}}$ and $f_{j}=\mathrm{i}\frac{p_{j}^{\dagger}-p_{j}}{\sqrt{2}}$. To describe a Gaussian state, it is convenient to use the first moment vector $\left\langle \vec{\xi} \right\rangle$ with $\vec{\xi}=(x_{U}, p_{U}, x_{L}, p_{L})$ and the second moment represented in the phase space by the covariance matrix $\Gamma$. For a Gaussian state \(\rho\), the elements of the covariance matrix are given by 
$\Gamma_{i, j}=\langle \frac{\Delta \xi_{i} \Delta \xi_{j}+\Delta \xi_{j} \Delta\xi_{i}}{2}\rangle$
with $\Delta \xi_{i}=\xi_{i}-\langle \xi_{i}\rangle$. Generally, the covariance matrix $\Gamma$ can be expressed in the following form \cite{RevModPhys.84.621},
\begin{equation} \label{GG-Cmatrix}
\Gamma=\left(\begin{array}{cc}
A & C^{\mathrm{T}} \\
C & B  
\end{array}\right),
\end{equation}
where matrices $A$, $B$ and $C$ satisfy the condition $A^{\mathrm{T}}=A$, $B^{\mathrm{T}}=B$ and $C^{\mathrm{T}}=C$, and they are real matrices. For simplification, one can define $I_{a}=\mathrm{det} [A]$, $I_{b}=\mathrm{det} [B]$, $I_{c}=\mathrm{det} [C]$, and $I_{ab}=\mathrm{det} [\Gamma]$. Furthermore, the quadratures $\vec{\xi}$ satisfy the relation $[\xi_{i}, \xi_{j}]=\mathrm{i} \Omega$, where the symplectic matrix is given by $ $ $\Omega=\oplus_{j}\left(\begin{matrix}
0& 1\\-1 & 0
\end{matrix}\right)$.
The positivity under the partial transpose (PPT) criterion, based on the covariance matrix, can be used to measure the separability of two subsystems \cite{PhysRevLett.84.2726}. This criterion can be quantified as  $\tilde{d}_{\pm}\geq\frac{1}{2}$, where the symplectic invariants are defined as $\tilde{d}_{\pm}=\sqrt{\frac{I_{a}+I_{b}-2 I_{c} \pm \sqrt{(I_{a}+I_{b}-2 I_{c})^2-4 I_{ab}}}{2}}$. For the continuous variable system, the entanglement can be measured by \cite{PhysRevLett.87.167904}
\begin{equation}
E_{N}=\max \lbrace 0, -\ln 2 \tilde{d}_{-} \rbrace.
\end{equation} 
A higher value of $E_N$ indicates stronger quantum entanglement.

Using Gaussian measurements, the EPR steering for arbitrary bipartite Gaussian states can be defined by Adesso et al. \cite{ PhysRevLett.114.060403, Kogias:15}
\begin{align} \label{Steering}
\mathcal{G}^{a \rightarrow b} & \equiv \max \left[0, \frac{1}{2} \ln \frac{I_{a}}{4 I_{ab}}\right], \\ \mathcal{G}^{b \rightarrow a} & \equiv \max \left[0, \frac{1}{2} \ln \frac{I_{b}}{4 I_{ab}}\right], \label{Steering-R}
\end{align} 
where $\mathcal{G}^{a \rightarrow b}(\mathcal{G}^{b \rightarrow a})$ denotes the mode $a(b)$ can steer the mode $b(a)$ by Gaussian measurement. The other bipartite Reid criteria \cite{PhysRevA.40.913} from inference variances has many similarities with Adesso criteria, and both criteria are sufficient and necessary for all-Gaussian scenarios. As for the genuine one-way EPR steering, one can refer to Ref. \cite{PhysRevLett.116.160403}.  The EPR steering is an asymmetric aspect of quantum nonlocality. It can be categorized into three distinct cases: (i) \textit{no-way steering} with $\mathcal{G}^{a \rightarrow b}=\mathcal{G}^{b \rightarrow a}=0 $ when $4 I_{ab}>I_{m}>I_{m^{\prime}}$; (ii) \textit{one-way steering} with $\mathcal{G}^{a \rightarrow b}>0$ and $\mathcal{G}^{a \rightarrow b}=0$ or $\mathcal{G}^{b \rightarrow a}>0$ and $\mathcal{G}^{a \rightarrow b}=0$ when $I_{m}>4 I_{ab}>I_{m^{\prime}}$; (iii) \textit{two-way steering} with $\mathcal{G}^{a \rightarrow b}>0$ and $\mathcal{G}^{b \rightarrow a}>0$ when $I_{m}>I_{m^{\prime}}>4 I_{ab}$. The indexes $m$, $m^{\prime}$ can take as $m, m^{\prime}=a, b$. 

To give an intuitive understanding, one can introduce the average occupations to analyze the directionality of steering Eqs. (\ref{Steering}) and (\ref{Steering-R}),
\begin{equation}
N_{m}=\frac{\left\langle \delta d_{m}^{2}\right\rangle+\left\langle \delta f_{m}^{2} \right\rangle-1}{2},
\end{equation}
where $\delta d_{m}^{2}$, $\delta f_{m}^{2}$ denote the diagonal elements of covariance matrix in the original bare representation. The second-order correlation quantities can be solved as $\left\langle a^{\dagger} a \right\rangle=\sum_{j}(|w_{j}|^{2}+ |y_{j}|^{2}) p_{j}^{\dagger} p_{j}+y_j^2)$, $\left\langle a a^{\dagger} \right\rangle=\sum_{j}(|w_{j}|^{2}+ |y_{j}|^{2}) p_{j}^{\dagger} p_{j}+w_j^2)$, $\left\langle a^{\dagger 2} \right\rangle =-\sum_{j} w_{j} y_{j}^{*}(2 p_{j}^{\dagger} p_{j}+1)$, $\left\langle a^{2} \right\rangle =-\sum_{j} w^{*}_{j} y_{j}(2 p_{j}^{\dagger} p_{j}+1)$, $\left\langle b^{\dagger} b \right\rangle=\sum_{j}(|x_{j}|^{2}+ |z_{j}|^{2}) p_{j}^{\dagger} p_{j}+z_j^2)$, $\left\langle b b^{\dagger} \right\rangle=\sum_{j}(|x_{j}|^{2}+ |x_{j}|^{2}) p_{j}^{\dagger} p_{j}+x_j^2)$, $\left\langle b^{\dagger 2} \right\rangle =-\sum_{j} x_{j} z_{j}^{*}(2 p_{j}^{\dagger} p_{j}+1)$, $\left\langle b^{2} \right\rangle =-\sum_{j} x^{*}_{j} z_{j}(2 p_{j}^{\dagger} p_{j}+1)$, $\left\langle b^{\dagger} a^{\dagger} \right\rangle =-\sum_{j}[(w_{j} z_{j}^{*}+x_{j} y_{j}^{*}) p_{j}^{\dagger} p_{j}+x_{j} y_{j}^{*}]$, $\left\langle a^{\dagger} b \right\rangle =\sum_{j}[(w_{j} x_{j}^{*}+z_{j} y_{j}^{*}) p_{j}^{\dagger} p_{j}+z_{j} y_{j}^{*}]$, $\left\langle a b^{\dagger} \right\rangle =\sum_{j}[(w_{j}^{*} x_{j}+z_{j}^{*} y_{j}) p_{j}^{\dagger} p_{j}+x_{j} w_{j}^{*}]$, and $\left\langle a b \right\rangle =-\sum_{j}[(w_{j}^{*} z_{j}+x_{j}^{*} y_{j}) p_{j}^{\dagger} p_{j}+z_{j} w_{j}^{*}]$. In this case, $\left\langle a^{\dagger 2} \right\rangle \neq 0$, $\left\langle a^{2} \right\rangle \neq 0$, $\left\langle b^{\dagger 2} \right\rangle \neq 0$, and $\left\langle b^{2} \right\rangle \neq 0$, so the average occupations can exactly reflect the EPR steering at zero temperature. This is because $ p_{j}^{\dagger} p_{j}$ is a constant in this case (cf. Appendix \ref{Appendix B}). In order to effectively describe the EPR steering except the ground-state case, the purity $\mu_{\alpha}$ can be introduced \cite{PhysRevLett.114.060403}. The two marginal purities can be defined as $\mu_{a/b}=\frac{1}{4 I_{a/b}}$, and the global purity $\mu_{ab}=\frac{1}{16 I_{ab}}$. According to the definition of the EPR steering as shown in Eqs. (\ref{Steering}) and (\ref{Steering-R}), one can employ the purities to discuss different cases of the EPR steering.

It is observed that from the master equation (\ref{eq:common_ms}), some second moments $\left\langle d_j^2\right\rangle$, $\left\langle f_j^2\right\rangle$, $\left\langle d_j f_j\right\rangle$ are nonzero, others are zero, and we can express terms of nonzero as $\left\langle d_j^2\right\rangle=\left\langle f_j^2\right\rangle=\frac{1+2 N(\omega_j)}{2}$ (cf. Appendix \ref{Appendix B}). For the current system, the covariance matrix at steady-state in the polariton basis can solved as the following form
\begin{equation} \label{Cov Matrix}
\Gamma_{p}=\left(\begin{array}{cccc}
a_{1} & 0 & 0 & 0 \\
0 & a_{1} & 0 & 0 \\
0 & 0 & b_{1} & 0 \\
0 & 0 & 0 & b_{1}
\end{array}\right),
\end{equation}
where $a_{1}=\left\langle d_U^2\right\rangle$, and $b_{1}=\left\langle d_L^2\right\rangle$.
Consequently, one can convert the covariance matrix presented in Eq. (\ref{Cov Matrix}) back to its original form, indicated as $\sigma$, by using the appropriate transformation,
\begin{align}
\bar{U}=\left(\begin{array}{cc}
\cos \theta \bar{U}_{1} & -\sin \theta \bar{U}_{2} \\
\sin \theta \bar{U}_{3} & \cos \theta \bar{U}_{4} 
\end{array}\right),
\end{align}
where $\bar{U}_{1}=\left(\begin{matrix}
g_{+}(\frac{\omega_{U}}{\omega_{a}})& 0\\0 & g_{-}(\frac{\omega_U}{\omega_a})
\end{matrix}\right)$, $\bar{U}_{2}=\left(\begin{matrix}
g_{+}(\frac{\omega_{U}}{\omega_{b}})& 0\\0 & g_{-}(\frac{\omega_U}{\omega_b})
\end{matrix}\right)$, $\bar{U}_{3}=\left(\begin{matrix} 
g_{+}(\frac{\omega_{L}}{\omega_{a}})& 0\\0 & g_{-}(\frac{\omega_L}{\omega_a})
\end{matrix}\right)$, and $\bar{U}_{4}=\left(\begin{matrix}
g_{+}(\frac{\omega_{L}}{\omega_{b}})& 0\\0 & g_{-}(\frac{\omega_L}{\omega_b})
\end{matrix}\right)$ with $g_{\pm}(x)=f_{+}(x) \pm f_{-}(x)$.
Correspondingly, the covariance matrix of the original modes can be expressed as 
\begin{align} \label{G-Cmatrix}
\Gamma_{o}=\left(\begin{array}{cc}
A_{\Gamma_{o}} & C^{\mathrm{T}}_{\Gamma_{o}} \\
C_{\Gamma_{o}} & B_{\Gamma_{o}}  
\end{array}\right),
\end{align}
where $A_{\Gamma_{o}}$, $B_{\Gamma_{o}}$, $C_{\Gamma_{o}}$ are $2\times2$ diagonal matrices. The corresponding diagonal elements of these matrices $A_{\Gamma_{o}}=\mathrm{diag}[\omega_{a}(\frac{a_{1} \cos^2 \theta}{\omega_{U}} +\frac{b_{1} \sin^2 \theta}{\omega_{L}}), \frac{a_1 \cos^2 \theta \omega_{U}+b_1 \sin^2 \theta \omega_{L}}{\omega_{a}}]$, $B_{\Gamma_{o}}=\mathrm{diag}[\omega_{b}(\frac{a_{1} \sin^2 \theta}{\omega_{U}} +\frac{b_{1} \cos^2 \theta}{\omega_{L}} ), \frac{a_1 \sin^2 \theta \omega_{U}+b_1 \cos^2 \theta \omega_{L}}{\omega_{b}}]$ and $C_{\Gamma_{o}}=\cos \theta \sin \theta \mathrm{diag}[\sqrt{\omega_{a} \omega_{b}} (\frac{b_1}{\omega_{L}}-\frac{a_1}{\omega_{U}}), \frac{\omega_L b_1-\omega_U a_1}{\sqrt{\omega_a \omega_b}}$. In order to analyze the mechanism of EPR steering, one can solve condition of the purities $\mu_a=\mu_b$, i. e. $\mathrm{det}(A_{\Gamma_{o}})-\mathrm{det}(B_{\Gamma_{o}})=0$. It can be solved as $(a_1^2-b_1^2)\cos 2\theta=0$, and it means that there exists three cases: $(a_1^2-b_1^2)=0$, or $\cos 2\theta=0$, or $(a_1^2-b_1^2)=0$ and $\cos 2\theta=0$. For $a_1^2=b_1^2$, it means that the average populations of polariton operators satisfy the condition $N(\omega_U)=N(\omega_L)$. In this case, the system must be in the ground state. As existence of the diamagnetic term, $(N(\omega_U)-N(\omega_L))\cos 2\theta$ also reflects the symmetry of subsystem and reservoir coupling as analyzed in Appendix \ref{Appendix D}.  
 \begin{figure}
\centering
\includegraphics[width=1.\columnwidth]{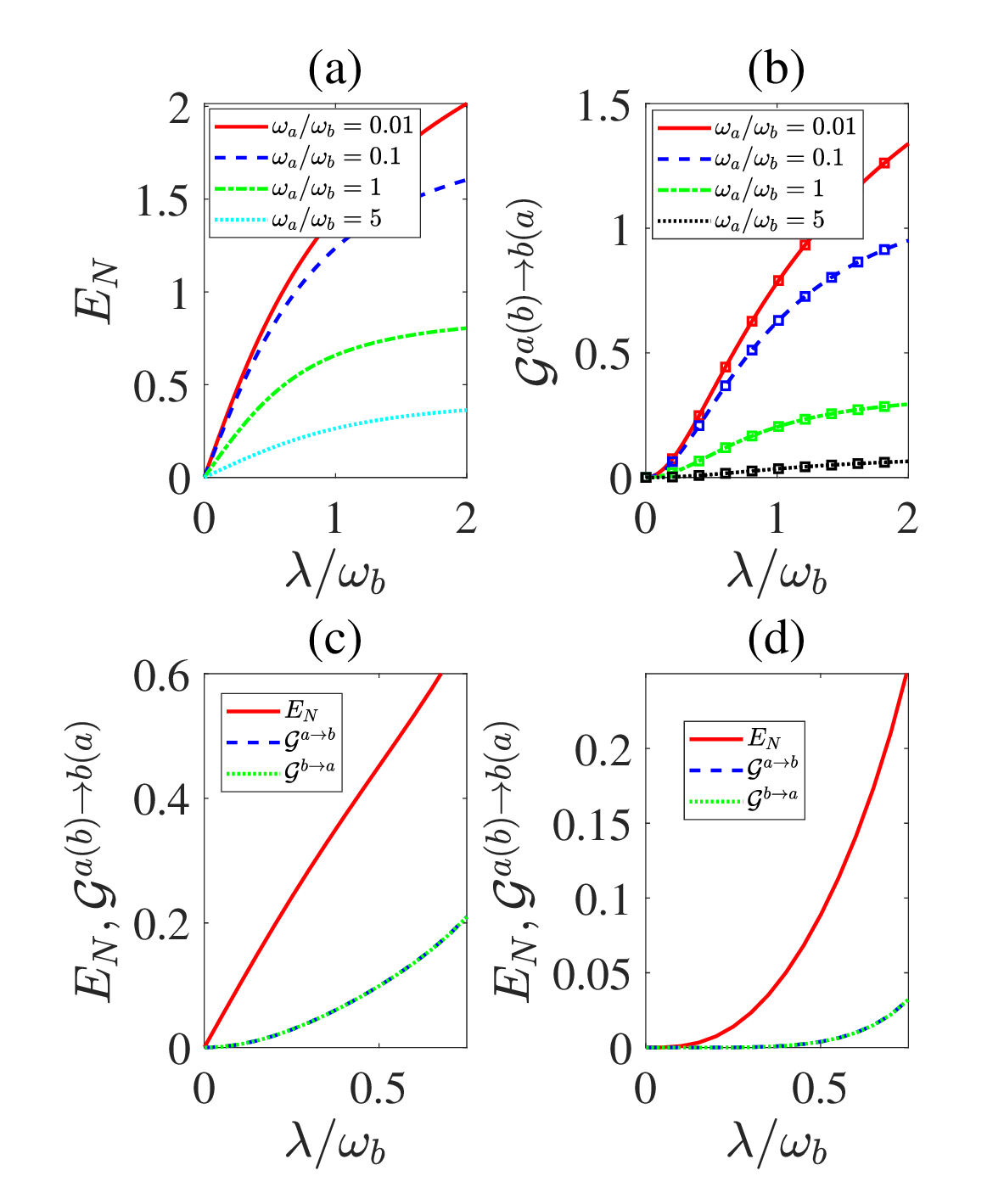}
\caption{\label{fig:En_ground}
The quantum entanglement is analyzed along with $\mathcal{G}^{a \rightarrow b}$, $\mathcal{G}^{b \rightarrow a}$ about the coupling strength $\lambda$ for different frequencies $\omega_{a}$. In panels (a) and (b), the internal coupling is considered full coupling, incorporating  
mode-mixing and mode-squeezing interactions. In contrast, panels (c) and (d) examine mode-squeezing ($\lambda=\lambda_2$, $\lambda_1=0$) and mode-mixing ($\lambda=\lambda_1$, $\lambda_2=0$) interactions separately, comparing them to the effect of full coupling. It is important to note that the square lines in (b) represent the reverse EPR steering $\mathcal{G}^{b \rightarrow a}$. The parameters in panels (c) and (d) can be set such that $\omega_{a} = \omega_{b}$.}
\end{figure}
 
The matrix elements of $\Gamma^{gs}_{o}$ in the ground state can be written in the following form,
$\Gamma^{gs}_{11}=\frac{\omega_a +\omega_b}{2\left(\omega_L+\omega_U\right)}$, $\Gamma^{gs}_{22}=\frac{\left(\omega_a+4 D+\omega_b \right)}{2 (\omega_L+\omega_U)}$, $\Gamma^{gs}_{33}=\frac{\left(\omega_a+4 D+\omega_b \right)}{2 (\omega_L+\omega_U)}$, $\Gamma^{gs}_{44}=\frac{\omega_a +\omega_b}{2\left(\omega_L+\omega_U\right)}$, $\Gamma^{gs}_{13}=-\frac{ \lambda }{\left(\omega_L+\omega_U\right)}$ and $\Gamma_{24}=\frac{\lambda}{\omega_L+\omega_U} $. Similarly, the elements of the covariance matrix $\Gamma^{\prime}_{o}$ for the non-zero temperature reservoir can be determined as 
\begin{align}
\begin{split}
&\Gamma^{\prime}_{11}=\frac{\omega_{a}(\coth(\frac{\omega_{L}}{2 T})\omega_{U}(\omega_{L}^2-\omega_{b}^2)-\coth(\frac{\omega_{U}}{2 T}) \omega_{L}(\omega_{U}^2-\omega_{b}^2))}{2 \omega_{L} \omega_{U}(\omega_{L}^2-\omega_{U}^2)},\\
&\Gamma^{\prime}_{22}=\frac{(\coth(\frac{\omega_{L}}{2 T})\omega_{L}(\omega_{L}^2-\omega_{b}^2)-\coth(\frac{\omega_{U}}{2 T}) \omega_{U}(\omega_{U}^2-\omega_{b}^2))}{2 \omega_{a}(\omega_{L}^2-\omega_{U}^2)}, \\
&\Gamma^{\prime}_{33}=\frac{\omega_{b}(\coth(\frac{\omega_{U}}{2 T}) \omega_{L}(\omega_{L}^2-\omega_{b}^2)-\coth(\frac{\omega_{L}}{2 T})\omega_{U}(\omega_{U}^2-\omega_{b}^2))}{2 \omega_{L} \omega_{U}(\omega_{L}^2-\omega_{U}^2)},\\
&\Gamma^{\prime}_{44}=\frac{(\coth(\frac{\omega_{U}}{2 T}) \omega_{U}(\omega_{L}^2-\omega_{b}^2)-\coth(\frac{\omega_{L}}{2 T})\omega_{L}(\omega_{U}^2-\omega_{b}^2))}{2 \omega_{b}(\omega_{L}^2-\omega_{U}^2)}, \\
&\Gamma^{\prime}_{13}=\frac{ \lambda\omega_{a} \omega_{b}(\coth(\frac{\omega_{L}}{2 T}) \omega_{U}-\coth(\frac{\omega_{U}}{2 T}) \omega_{L})}{\omega_{L} \omega_{U} (\omega^2_{L}-\omega^2_{U})},\\
&\Gamma^{\prime}_{24}=\frac{\lambda (\coth(\frac{\omega_{L}}{2T}) \omega_{L}-\coth(\frac{\omega_{U}}{2T}) \omega_{U})}{\omega_{L}^2-\omega_{U}^2},
\end{split}
\end{align}
and $\Gamma_{31}^{\prime}=\Gamma_{13}^{\prime}$, $\Gamma_{42}^{\prime}=\Gamma_{24}^{\prime}$.

\section{Results and discussion} \label{Sec. III}
\begin{figure}
\centering
\subfigure{\includegraphics[width=0.49\columnwidth]{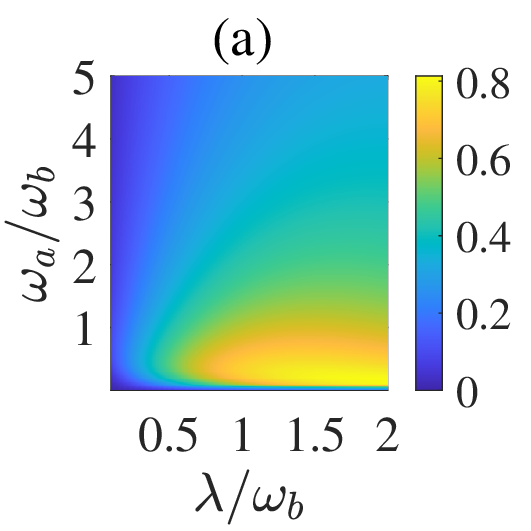}}
\subfigure{\includegraphics[width=0.49\columnwidth]{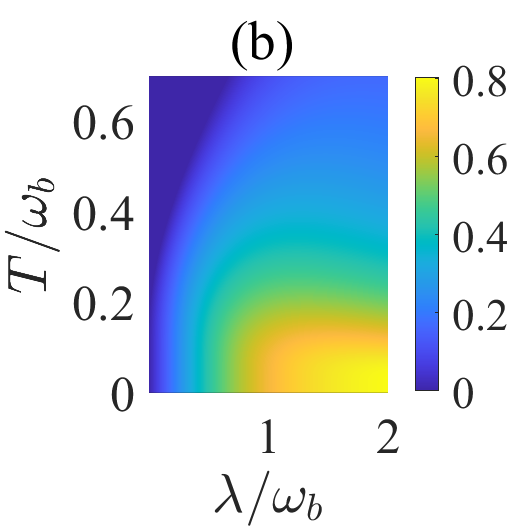}}
\caption{\label{fig:EN_thermal}
(a) Plot of quantum entanglement as a function of optical frequency $\omega_{a}$ and coupling strength $\lambda$; (b) Plot of quantum entanglement as a function of temperature $T$ and coupling strength $\lambda$. The parameter values are specified as follows: (a) $T = 0.15\omega_{b}$; (b) $\omega_{a} = \omega_{b}$. }
\end{figure}

We begin by examining bipartite quantum entanglement and EPR steering in the ground state, specifically highlighting the influence of mode-mixing and squeezing interactions on quantum correlations. It is demonstrated that the non-diagonal elements of the covariance matrix are zero, indicating no entanglement when $\lambda\rightarrow 0$. In the zero-temperature regime $T\rightarrow0$, the expression for quantum entanglement can be written as $E_{N}=-\ln (2\tilde{d}_{-})=-\ln (\frac{1}{\omega_{L}+\omega_{U}}\sqrt{\frac{ (2\lambda \sqrt{\omega_{a} \omega_{b}} -\sqrt{\zeta})^2}{\omega_{U} \omega_{L}}}$ with $\zeta=(4 D \omega_{a}+\omega_{a}^2+\omega_{L} \omega_{U})(\omega_{b}^2+\omega_{L} \omega_{U})$. As shown in Fig. \ref{fig:En_ground}(a), the entanglement can be enhanced by increasing the internal coupling strength; particularly in deep strong coupling regime, larger entanglement can be achieved.
Additionally, a lower ratio of $\omega_{a}/\omega_{b}$ allows for a larger range of entanglement.
\begin{figure}
\centering
\includegraphics[width=0.75\columnwidth]{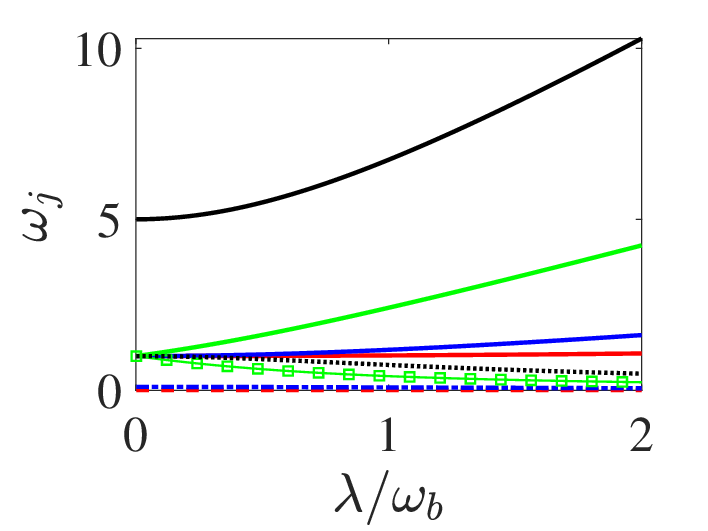}
\caption{\label{fig:omega_j}
The polaritons frequencies $\omega_{j}$ with $j=U, L$ are plotted against the coupling strength $\lambda$ for various optical frequencies $\omega_{a}$. The red, blue, green, and black lines represent the ratios $\omega_{a}/\omega_{b} = 0.01, 0.1, 1, 5$, respectively. Solid lines indicate upper polaritons, whereas other lines represent lower polaritons.}
\end{figure}
Using Eqs. (\ref{Steering}) and (\ref{Steering-R}), we derive the expression for EPR steering as $\mathcal{G}^{a \rightarrow b}=\mathcal{G}^{b \rightarrow a}=\frac{1}{2} \ln \frac{\omega_L \omega_U (\omega_L+\omega_U)^2 \zeta}{(-4 \lambda^2 \omega_a \omega_b+\zeta)^2}$. From Fig. \ref{fig:En_ground}(b), we can see that the trend of steering closely resembles that of quantum entanglement, as indicated by the condition $\mathcal{G}^{a \rightarrow b}=\mathcal{G}^{b \rightarrow a}>0$. This suggests the existence of a two-way regime of EPR steering. Analyzing the covariance matrix $\sigma$ in the ground state reveals that $N_{a}=N_{b}$, leading to the symmetry of steering. To offer a clearer understanding, we plot the quantum entanglement and EPR steering against the mode-squeezing coupling shown in Fig. \ref{fig:En_ground}(c) and mode-mixing coupling shown in Fig. \ref{fig:En_ground}(d) for the resonant case $\omega_a=\omega_b$. The non-resonant case displays similar behavior, but we will not elaborate on it here.
It is noteworthy that mode-mixing coupling has a negligible impact on the production of quantum entanglement and EPR steering when the coupling roughly satisfies the condition $\lambda/\omega_{b}\leq 0.2$, with the squeezing coupling being the dominant factor. However, as the internal coupling increases beyond this threshold, mode-mixing coupling begins to take on a more significant role. In this regime, the enhanced quantum correlation results from the synergistic effects of mode-mixing and mode-squeezing couplings. This can be seen from the view of quantum optics. As shown in Eq.(\ref{H_{SS}}), the generation of quantum entanglement stems from the squeezed-type coupling $(a^{\dagger} b^{\dagger}+a b)$, which becomes important in the ultrastrong and deep strong coupling regimes. However, the beam-splitter term $(a^{\dagger} b+a b^{\dagger})$ does bot produce any entanglement.
The ground state of the light-matter system can read as $\left| \Phi_{gs}\right\rangle=\left| 0_{U} \right\rangle \left| 0_{L} \right\rangle$ as the polariton operators $p_{\pm}$ can annihilate it, i.e. $p_{\pm}\left| \Phi_{gs}\right\rangle=0$.  The linear polariton operators are linear superposition of bare operators, the ground state can be a squeezed state. Hence the entanglement can be enhanced in the combined effect of squeezed-type and beam-splitter couplings.

In what follows, we will explore the generation and manipulation of one-way EPR steering in coupled systems that share a common heat reservoir. As illustrated in Fig. \ref{fig:EN_thermal}(a), the lower $\omega_{a}/\omega_{b}$ and ultrastrong and deep strong coupling regimes can produce the maximal quantum entanglement. Let us now consider the thermal effect on quantum entanglement as shown in Fig. \ref{fig:EN_thermal}(b). We find that the thermal noise is detrimental to the survival of quantum entanglement, which is expected. To understand the effect of coupling strength and optical frequency on entanglement, we plot the polariton frequencies $\omega_{j}$ as a function of coupling strength $\lambda$ for different $\omega_{a}$. As shown in Fig. \ref{fig:omega_j}, in the limit $\lambda\rightarrow 0$, the two polariton frequencies converge to $\omega_{U}\rightarrow \omega_{a}$ and $\omega_{L}\rightarrow \omega_{b}$. We also find that the polariton gap increases with greater coupling strength $\lambda$ while keeping $\omega_{a}$. A lower ratio of $\omega_{a}/\omega_{b}$ results in a smaller polariton gap for a fixed coupling strength $\lambda$. In the ultrastrong and deep strong coupling regimes, the energy gap between the two branches primarily originates from the lower polariton, which plays a critical role in quantum entanglement, as demonstrated in Fig. \ref{fig:EN_thermal}(a).

\begin{figure}
\centering
\subfigure{\includegraphics[width=1.\columnwidth]{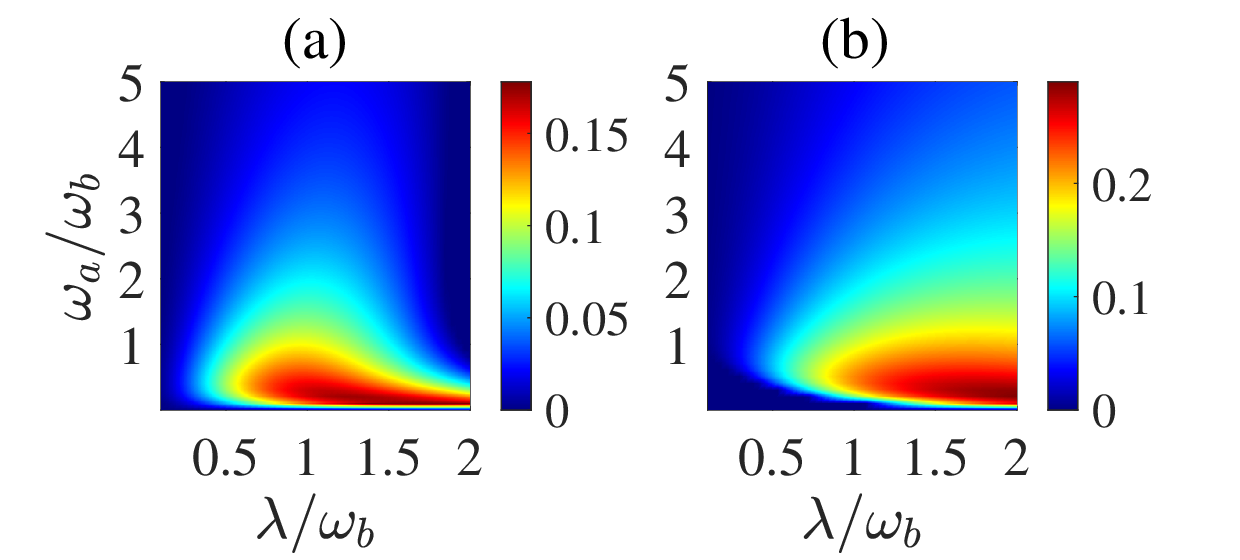}}
\subfigure{\includegraphics[width=1.\columnwidth]{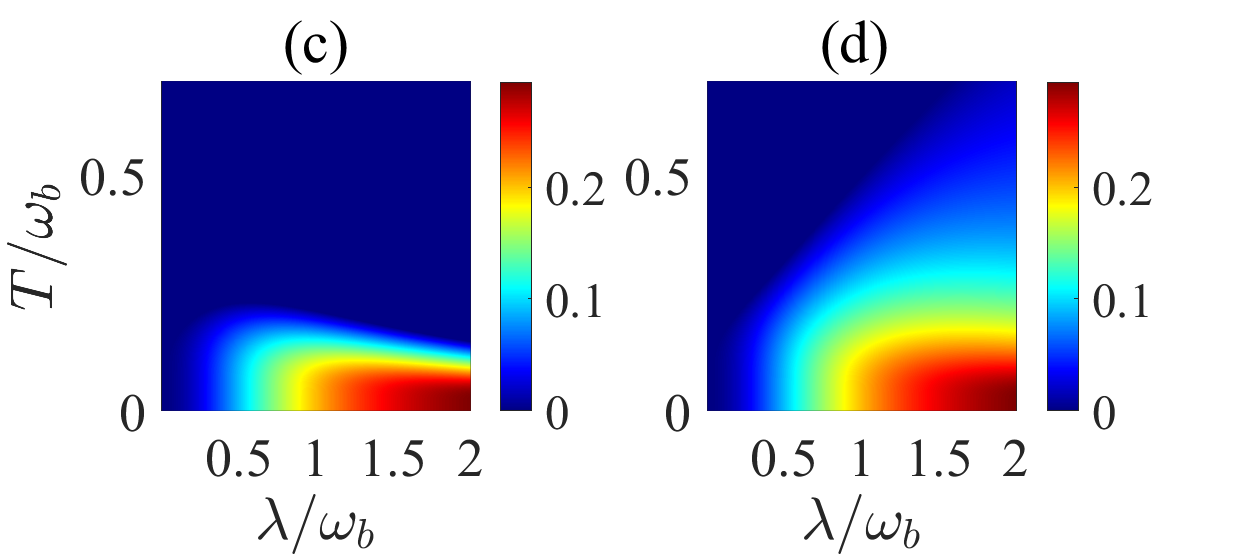}}
\caption{\label{fig:steering_thermal}
Quantum steering $\mathcal{G}^{a \rightarrow b}$ (a) and $\mathcal{G}^{b \rightarrow a}$ (b) are analyzed in relation to the frequency $\omega_{a}$ and the coupling strength $\lambda$; for panels (c) and (d), $\mathcal{G}^{a \rightarrow b}$ and $\mathcal{G}^{b \rightarrow a}$ are examined as functions of temperature $T$ and $\lambda$. The parameters are set as follows: for panels (a) and (b), $T=0.15\omega_{b}$; and for panels (c) and (d), $\omega_{a}=\omega_{b}$. }
\end{figure}
\begin{figure}[htbp]
\centering
\includegraphics[width=1.2\columnwidth]{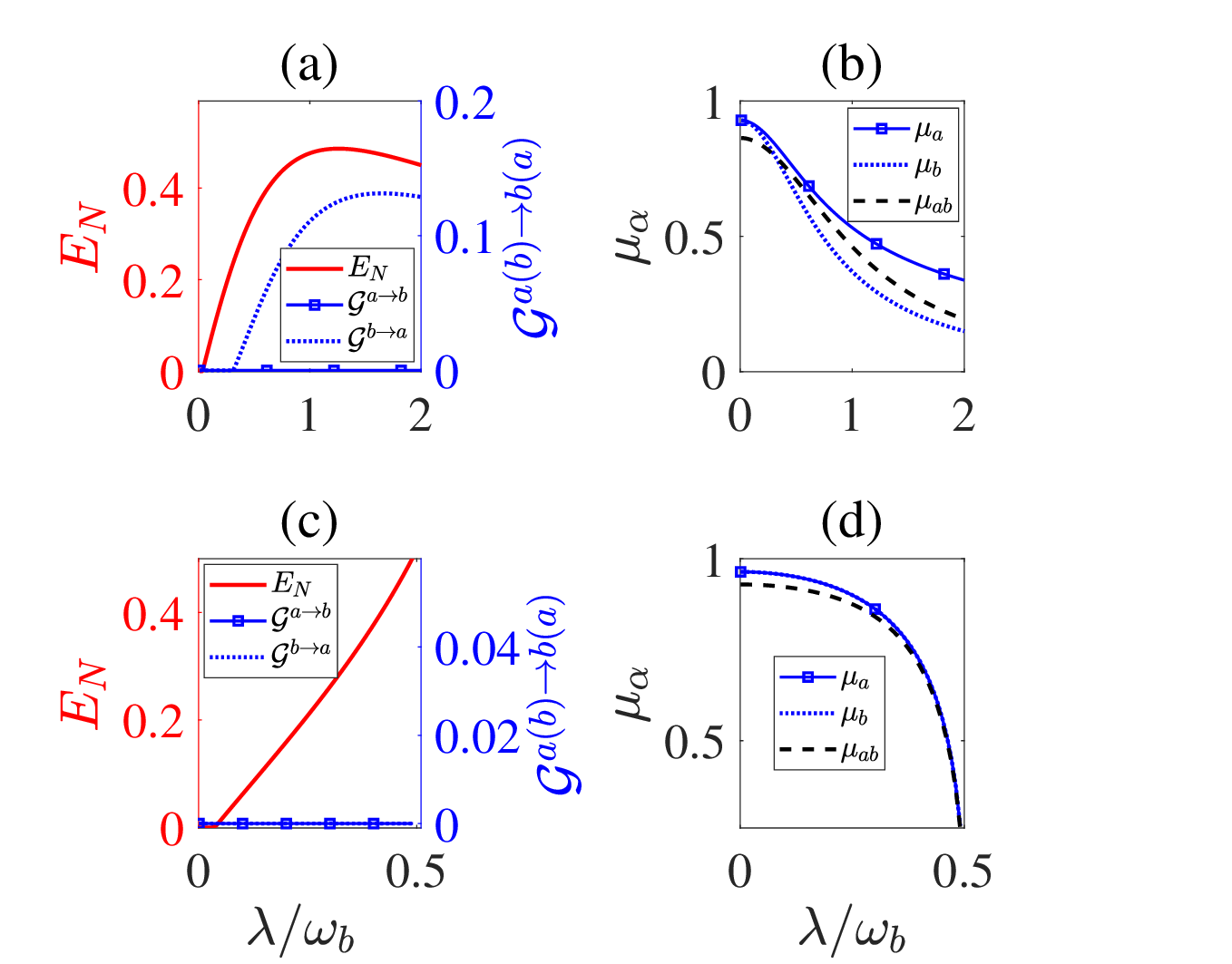}
\caption{\label{fig:steering_res}
In panels (a) and (c), the plot illustrates quantum entanglement and EPR steering $G^{a \rightarrow b}$ and $G^{b \rightarrow a}$ versus coupling strength $\lambda$. We also consider the case that the Hamiltonian Eq. (\ref{H_{S}}) without the diamagnetic term compared to the Hopfield model. In panels (b) and (d), the purities correspond to the values in panels (a) and (c). The parameters are $\omega_a = \omega_b$, and $T=0.25 \omega_b$.}
\end{figure}

EPR steering plays a crucial role in one-sided device-independent quantum key distribution. In Figs. \ref{fig:steering_thermal} (a) and \ref{fig:steering_thermal} (b), it is observed that the one-way EPR steering can be implemented for wide optical frequencies $\omega_a$ range in the ultrastrong and deep strong coupling regimes. Figs. \ref{fig:steering_thermal} (c) and \ref{fig:steering_thermal} (d) illustrate that in moderate temperature ranges, along with ultrastrong and deep strong coupling regimes, one-way EPR steering can indeed be achieved. As shown in Fig. \ref{fig:steering_res} (a), the one-way EPR steering $\mathcal{G}^{b \rightarrow a}$ for a resonant case can be realized in the ultrastrong and deep strong coupling regimes, whereas $\mathcal{G}^{a \rightarrow b}$ does not exist in the system. This can be confirmed in Fig. \ref{fig:steering_res} (b), where the purities condition $\mu_{b}<\mu_{ab}<\mu_{a}$ indicates the steering from mode $b$ to $a$ mode. To understand the mechanism of one-way EPR steering, we consider the case $D=0$ and plot the quantum entanglement, EPR steering, as a function of the coupling strength shown in Fig. \ref{fig:steering_res} (c). From Fig. \ref{fig:steering_res} (c), there is no EPR steering. This can be seen from Fig. \ref{fig:steering_res} (d), the purities $\mu_{a}, \mu_{b} \geqslant \mu_{ab}$, and the related analysis also refers to Appendix \ref{Appendix C}. In this case, it means that the system has no asymmetry. One considers the nonresonant $\omega_a \neq \omega_b$, and the one-way EPR steering may be realized as shown in Fig. \ref{fig8}. Here, we don't emphasize it. Compared to the case $D=0$, we can find that the diamagnetic term for a resonant case can produce one-way EPR steering, particularly in the ultrastrong and deep strong coupling regimes. In order to give a intuitive understanding, we express the global master equation Eq. (\ref{eq:common_ms}) in the local presentation as shown in Appendix \ref{Appendix D} and find that the diamagnetic term lead to the asymmetry of subsystem-reservoir coupling, which induces the one-way EPR steering. This is different from previous proposals \cite{WOS:000308695100011, WOS:000349934700022, PhysRevA.95.052114, PhysRevLett.119.160501, PhysRevA.98.042115, PhysRevApplied.15.024042}.

\begin{figure}[htbp]
\centering
\includegraphics[width=1.2\columnwidth]{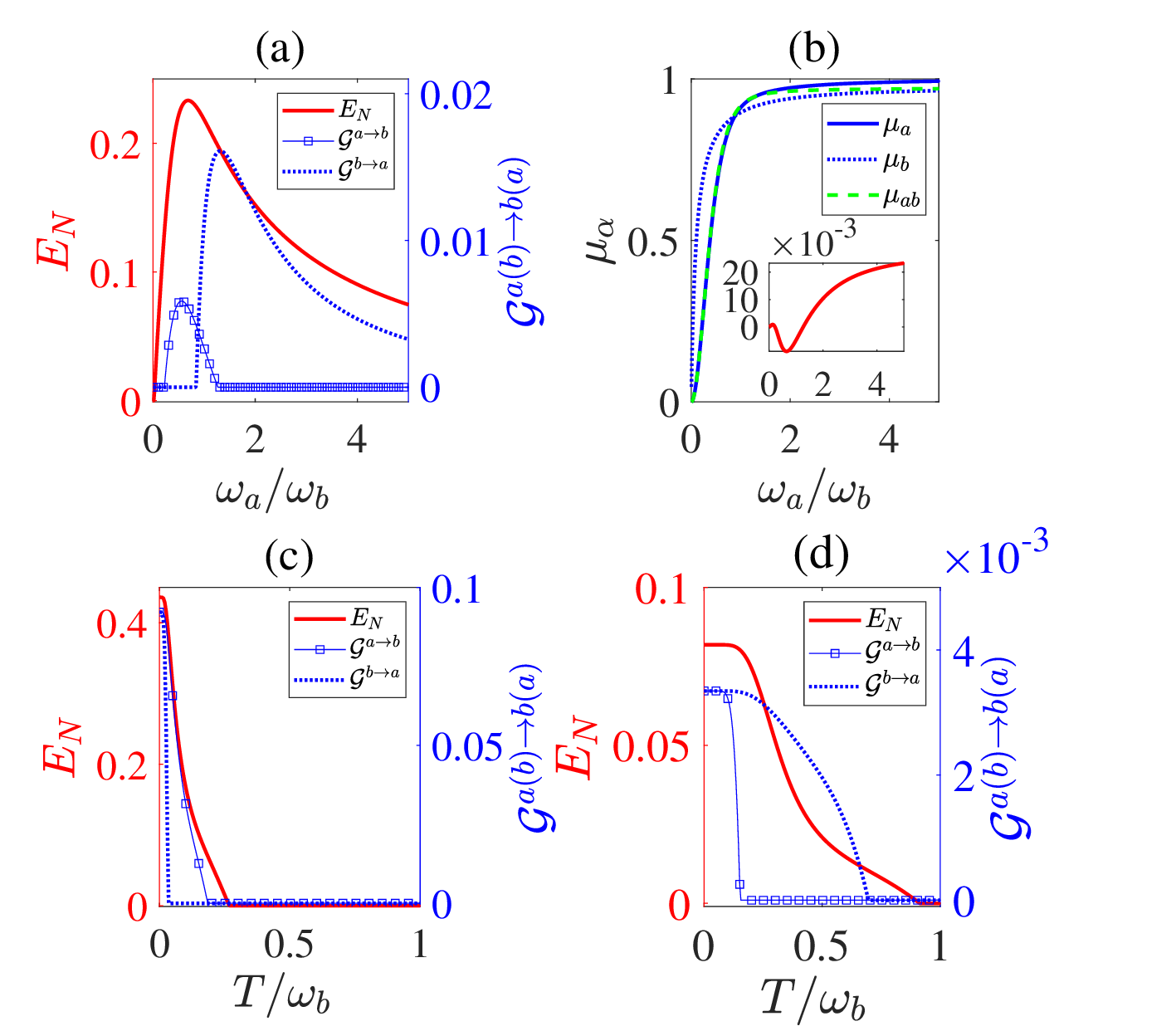}
\caption{\label{fig:steering_off_res}
The plot illustrates quantum entanglement and steering $G^{a \rightarrow b}$ and $G^{b \rightarrow a}$ versus frequency $\omega_a$ in panel (a) and temperature $T$ in panel (c). In panel (b), the purities correspond to the values in panel (a), and the inset view shows the bias of purities $\mu_a-\mu_{ab}$ for comparison. The parameters for panels (a) and (b) are set to $T/\omega_{b} = 0.2$. The optical frequencies can take $\omega_a=0.1\omega_b$ for panel (c), and $\omega_a=5 \omega_b$ for panel (d), as well as the coupling strengths is $\lambda/\omega_b=0.25$ for all panels.}
\end{figure}
Further, we also discuss the nonresonant case as demonstrated in Fig. \ref{fig:steering_off_res}. From Fig. \ref{fig:steering_off_res} (a), we can find that both one-way EPR steering $\mathcal{G}^{a \rightarrow b}$ and $\mathcal{G}^{b \rightarrow a}$ can be realized for moderate the optical frequencies $\omega_a$. It mainly originates from the asymmetry with the combined effect of off-resonance condition and the existence of a diamagnetic term. In the same way, it can be evident from the purities $\mu_a$, $\mu_b$, and $\mu_{ab}$ as shown in Fig. \ref{fig:steering_off_res} (b). In low frequencies $\omega_a$, the purities satisfy the condition $\mu_b>\mu_a>\mu_{ab}$, there is no steering. As the frequencies gradually increases, the purities become $\mu_a<\mu_{ab}<\mu_{b}$, the one-way EPR steering $\mathcal{G}^{a \rightarrow b}>0$ can be generated. When $\mu_a=\mu_b$, one can obtain solution of the optical frequency $\omega_a=\frac{-4 \lambda^2+\sqrt{16 \lambda^4+4 \omega_b^2}}{2 \omega_b}=0.8828$ by solving the equation $\cos 2\theta=0$. It is shown that $\omega_a>0.8828$ for $\mu_b<\mu_{ab}<\mu_{a}$ with $\mathcal{G}^{b \rightarrow a}>0$ corresponding to one-way steering. As shown in Figs. \ref{fig:steering_off_res} (c) and \ref{fig:steering_off_res} (d), there are three scenarios: two-way EPR steering, one-way EPR steering, and no-way EPR steering. It has also been shown that one-way EPR steering becomes feasible at a larger temperature range. Additionally, $E_N$ can be implemented with a wide temperature area, as demonstrated in Figs. \ref{fig:steering_off_res} (c) and \ref{fig:steering_off_res} (d).

We would also like to discuss the experimental proposals for the model. Many physical systems can be engineered to realize the Hopfield model, which consists of two coupling oscillators with diamagnetic terms. In particular, Ref. \cite{PhysRevApplied.20.024039} shows a viable way to obtain the ultrastrong coupling regime with a yttrium iron garnets (YIG) film coupled with superconducting coplanar waveguide (CPW) resonator based on coupled superconductor-ferrimagnet hybrid architectures. The experimental parameters can take $\omega_{a}/2 \pi=8.65 \mathrm{GHz}$ for the cavity frequency, $\lambda/2 \pi=2 \mathrm{GHz}$ for the coupling strength, and $C=5 \times 10^{4}$ for the cooperativity. All other system parameters are scaled with cavity frequency. Moreover, as shown in Ref. \cite{baranov2020ultrastrong}, the frequencies of nanoparticle plasmon and photons can take a few $\mathrm{eV}$, for example, $\omega_{a}=\omega_{b}=1 \mathrm{eV}$ for resonant case. The value of the interaction strength can reach the ultrastrong coupling regime $\lambda/\omega_a=0.55$. The superconducting circuits can also realized this model \cite{nataf2010no}. In addition, the deep strong coupling regime of light and matter can be realized in Refs.\cite{mornhinweg2024mode, halbhuber2020non}. In a real experimental scenario, the system is unavoidably coupled to an external reservoir environment. The heat reservoir can be modelled by a general noiseless resistor with a fluctuating voltage source \cite{RevModPhys.82.1155, Cattaneo}.

From an experimentally accessible view, the system's dynamics can be described by the covariance matrix's first and second moments, and the related quadrature components can be measured, which can be reconstructed by the tomographic reconstruction \cite{PhysRevLett.118.230501, PhysRevLett.102.020502, barzanjeh2019stationary, Laurat_2005}. As shown in Ref. \cite{Laurat_2005}, one can measure the orthogonal quadratures of light mode to precisely estimate the entries of the covariance matrix of bosonic modes. Hence the Adesso criterion can be experimentally accessible and one can follow the relevant experiment demonstrations of quantum entanglement and EPR steering in Refs. \cite{barzanjeh2019stationary, chen2020entanglement, kotler2021direct, palomaki2013entangling, PhysRevLett.121.100401, PhysRevLett.121.103602, PRXQuantum.3.030102}. In addition, one can also employ the Reid criterion to measure quadratures components by employing the homodyne detections. Based on these researches, the experimental implementation of our scheme can be further processed.


\section{Conclusion}
\label{conclusion}

In conclusion, we have proposed an efficient scheme for engineering quantum entanglement and EPR steering within a Hopfield model comprising ultrastrongly coupled oscillators interacting with a common heat reservoir. We start by examining the scenario of two coupled oscillators in their ground state. Our results demonstrate that the squeezing interaction is crucial in establishing quantum correlations in the weak or normal strong coupling regimes. As the coupling strength increases, particularly in the ultrastrong-coupling regime, we find that quantum correlations emerge from the combined effects of mode-mixing and mode-squeezing interaction. Moreover, we also explore the thermal impact on quantum entanglement and EPR steering. The one-way EPR steering can be feasible for resonant cases in the ultrastrong and deep strong coupling regimes induced by the diamagnetic term. Further, one-way EPR steering can also realized for nonresonant cases. In addition, our analysis shows that steady-state quantum entanglement and one-way EPR steering are robust to variations in the temperature of the heat reservoir and the optical frequency. Our results can also be extended to multi-mode cavities \cite{PhysRevB.72.115303} and may pave the way for new explorations in quantum information processing based on quantum correlations.

\section*{Acknowledgement}
This work was supported by , and the China Postdoctoral Science Foundation under Grant No. 2024M760829, and the National Natural Science Foundation of China under Grant No. 12175029.

\appendix
\section{The Hopfield transformation} \label{Appendix A}

Based on the commutation relation $[p_{j}, H_{S}]=\omega_{j} p_{j}$, the polariton frequencies $\omega_{j}$ of the system's Hamiltonian can be solved by diagonalizing the following equation,
\begin{equation} \label{eig-Eq}
\left(\begin{array}{cccc}
\omega_{a}+2 D & \lambda & 2 D & \lambda \\
\lambda & \omega_{b} & \lambda & 0 \\
-2D & -\lambda & -(\omega_{a}+2D) & -\lambda \\
-\lambda & 0 & -\lambda & -\omega_{b}
\end{array}\right).
\end{equation}
The eigenstates associated with these frequencies can be represented using a transformation matrix $S$. 
By utilizing this unitary transformation $S$, one can convert the original model's creation and annihilation operators into the polaritonic modes $\left(p_U, p_L, p_U^{\dagger}, p_L^{\dagger}\right)$, i.e., $\left(p_U, p_L, p_U^{\dagger}, p_L^{\dagger}\right)^{\mathrm{T}}=S \left(a, b, a^{\dagger}, b^{\dagger}\right)^{\mathrm{T}}$. 
The elements of the transformation matrix $S$ can be detailed as \cite{PhysRevA.91.063840}
\begin{align} \label{Eq.coefficients-1}
\left(\begin{array}{l}
w_{U} \\
x_{U} \\
y_{U} \\
z_{U}
\end{array}\right)=
\left(\begin{array}{c}
\cos \theta f_{+}\left(\frac{\omega_U}{\omega_a}\right) \\
-\sin \theta f_{+}\left(\frac{\omega_U}{\omega_b}\right) \\
\cos \theta f_{-}\left(\frac{\omega_U}{\omega_a}\right) \\
-\sin \theta f_{-}\left(\frac{\omega_U}{\omega_b}\right)
\end{array}\right),
\end{align}
and 
\begin{align} \label{Eq.coefficients-2}
 \left(\begin{array}{l}
w_{L} \\
x_{L} \\
y_{L} \\
z_{L}
\end{array}\right)=
\left(\begin{array}{c}
\sin \theta f_{+}\left(\frac{\omega_L}{\omega_a}\right) \\ \cos \theta f_{+}\left(\frac{\omega_L}{\omega_b}\right) \\ \sin \theta f_{-}\left(\frac{\omega_L}{\omega_a}\right) \\ \cos \theta f_{-}\left(\frac{\omega_L}{\omega_b}\right) 
\end{array}\right),
 \end{align}
 where $\cos 2 \theta=\frac{\omega_a^2+4 D \omega_a-\omega_b^2}{\omega_U^2-\omega_L^2}$, $\sin 2 \theta=-\frac{4 \lambda \sqrt{\omega_a \omega_b}}{\omega_U^2-\omega_L^2}$ with $\theta<0$ and
$f_{ \pm}(x)=\frac{1}{2}\left(\sqrt{x} \pm \frac{1}{\sqrt{x}}\right)$.

\section{Covariance matrix formalism} \label{Appendix B}
  The evolution of the second moments for a reduced system interacting with a common reservoir can be expressed as
\begin{align}
\begin{split}
\frac{\mathrm{d} p_{j}^{\dagger 2}}{\mathrm{d} t}&= \Upsilon^{C }_{j} p^{\dagger 2}_{j}+2 \mathrm{i} \omega_{j} p_{j}^{\dagger 2},\\
\frac{\mathrm{d} p_{j}^2}{\mathrm{d} t}&=\Upsilon^{C}_{j}p^{2}_{j}-2 \mathrm{i} \omega_{j} p_{j}^{\dagger 2},\\
\frac{\mathrm{d} p_{j}^{\dagger} p_{j}}{\mathrm{d} t}&=[-\Gamma^{C }(-\omega_{j}) p^{\dagger}_{j} p_{j}+\Gamma^{C }(\omega_{j}) p_{j} p_{j}^{\dagger}],\\
\frac{\mathrm{d} p_{j} p^{\dagger}_{j}}{\mathrm{d} t}&=[-\Gamma^{C }(-\omega_{j}) p^{\dagger}_{j} p_{j}+\Gamma^{C }(\omega_{j}) p_{j} p_{j}^{\dagger}],\\
\frac{\mathrm{d} p^{\dagger}_{U} p_{L}}{\mathrm{d} t}&=[\mathrm{i}(\omega_{U}-\omega_{L})+\sum_{j}\frac{\Upsilon^{C }_{j}}{2}]p^{\dagger}_{U} p_{L},
\end{split}
\end{align}
where $\Upsilon^{C}_{j}=[-\Gamma^{C}(-\omega_{j})+\Gamma^{C}(\omega_{j})]$. Thus, one can derive the second moments from these equations to obtain the covariance matrix presented in Eq. (\ref{Cov Matrix}).

\section{The Hopfield model without $A^2$ term} \label{Appendix C}

\begin{figure}[htbp]
\centering
\includegraphics[width=0.7\columnwidth]{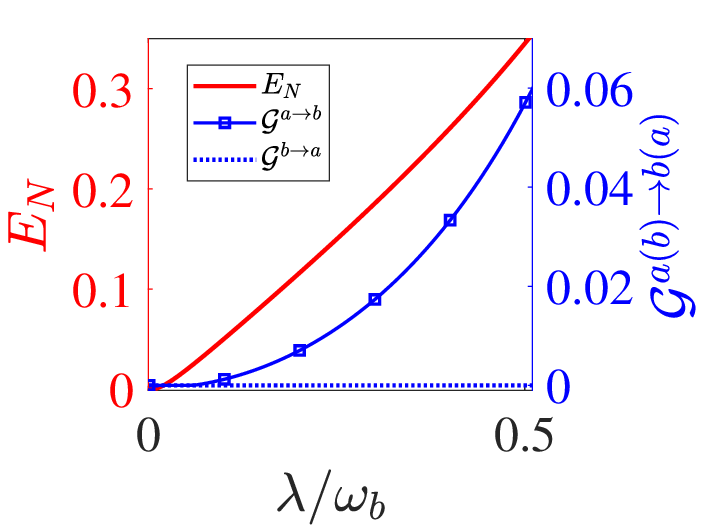}
\caption{\label{fig8}
The plot illustrates quantum entanglement and steering $G^{a \rightarrow b}$ and $G^{b \rightarrow a}$ versus the coupling strength $\lambda$. The parameters can take $\omega_a=2 \omega_b$, and other parameters are the same as in Fig. \ref{fig:steering_res}. }
\end{figure}

  The related coefficients of diagonalization of the system's Hamiltonian can be solved as 
$w_{j}=-\frac{(\omega_a+\omega_j)(\omega_b+\omega_j)}{2 \mathcal{N}_{j} \lambda \omega_a}$, $x_j=\frac{1}{\mathcal{N}_{j}}(-1+\frac{(\omega_b+\omega_j)(\omega_a^2-\omega_j^2)}{2 \lambda^2 \omega_a})$, $y_j=-\frac{(\omega_a-\omega_j)(\omega_b+\omega_j)}{2 \mathcal{N}_{j} \lambda \omega_a}$, and $z_{j}=\frac{1}{\mathcal{N}_{j}}$ with $\mathcal{N}_{j}=\frac{\sqrt{(\omega_b+\omega_j)(-4 \lambda^2\omega_a^3+(\omega_b+\omega_j)(\omega_a^2-\omega_j)^2+4 \lambda^2 \omega_a \omega_j (\omega_b+2 \omega_j))}}{2 g^2 \omega_a}$.
\begin{widetext}
 In this case, the covariance matrix $G$ for original modes can be expressed as the following form,
\begin{align}
\begin{split}
&G_{11}=\frac{\lambda^2 (\mathcal{N}_{U}^2(1+2 N(\omega_{U}))(\omega_b+\omega_L)^2(\omega_a^2-\omega_L^2)^2+\mathcal{N}_{L}^2(1+2 N(\omega_{L}))(\omega_b+\omega_U)^2(\omega_a^2-\omega_U^2)^2)}{2 (\omega_b+\omega_L)^2(\omega_b+\omega_U)^2(\omega_L^2-\omega_U^2)^2},\\
&G_{22}=\frac{\lambda^2 \omega_a^2(\mathcal{N}_{U}^2(1+2N(\omega_U))\delta(\omega_L)+\mathcal{N}_{L}^2(1+2 N(\omega(\omega_L))\delta(\omega_U))}{2(\omega_L-\omega_U)^2((\omega_b+\omega_L)(\omega_b+\omega_U)(\omega_a^2+\omega_L \omega_U)-4 \lambda^2 \omega_a (\omega_b+\omega_L+\omega_U))^2)},\\
&G_{33}=\frac{2 \lambda^4 \omega_a^2(\mathcal{N}_{U}^2 (1+2N(\omega_U))(\omega_b+\omega_L)^2+\mathcal{N}_{L}^2 (1+2 N(\omega_L)) (\omega_b+\omega_U)^2)}{(\omega_b+\omega_L)^2(\omega_b+\omega_U)^2(\omega_L^2-\omega_U^2)^2},\\
&G_{44}=\frac{2 \lambda^4(\mathcal{N}_{U}^2(1+2 N(\omega_{U})) \omega_L^2(\omega_b+\omega_L)^2+\mathcal{N}_{L}^2(1+2 N(\omega_L)) \omega_U^2(\omega_b+\omega_U)^2)}{(\omega_L-\omega_U)^2((\omega_b+\omega_L)(\omega_b+\omega_U)(\omega_a^2+\omega_L \omega_U)-4 \lambda^2 \omega_a (\omega_b+\omega_L+\omega_U))^2)},\\
&G_{13}=\frac{g^3 \omega_a(\mathcal{N}_{U}^2(1+2 N(\omega_U))(\omega_b+\omega_L)^2(\omega_a^2-\omega_L^2)+\mathcal{N}_{L}^2(1+2N(\omega_L))(\omega_b+\omega_U)^2(\omega_a^2-\omega_U^2))}{(\omega_b+\omega_L)^2(\omega_b+\omega_U)^2(\omega_L^2-\omega_U^2)^2},\\
&G_{24}=\frac{\lambda^3 \omega_a^2(\mathcal{N}_{U}^2(1+2N(\omega_U)) \xi(\omega_L)+\mathcal{N}_{L}^2(1+2N(\omega_L)) \xi(\omega_U))}{(\omega_L-\omega_U)^2((\omega_b+\omega_L)(\omega_b+\omega_U)(\omega_a^2+\omega_L \omega_U)-4 \lambda^2 \omega_a (\omega_b+\omega_L+\omega_U))^2)},
\end{split}
\end{align}
where $\delta(\omega_j)=(4 \lambda^2\omega_a-\omega_a^2(\omega_b+\omega_j)+\omega_j^2(\omega_b+\omega_j))^2$, $\xi(\omega_j)=\omega_j (\omega_b+\omega_j)(-4\lambda^2\omega_a+(\omega_b+\omega_j)(\omega_a^2-\omega_j^2))$, $G_{31}=G_{13}$, and $G_{42}=G_{24}$. 
\end{widetext}
For resonant case, the elements of covariance matrix can reduce to $G_{11}=G_{33}= \lambda^2\frac{\sum_{j}\mathcal{N}_{j}^2(1+2 N(\omega_{j}))}{8(\omega_b+\sqrt{\omega_b^2+2 \lambda \omega_b})^2}$, $G_{22}=G_{44}=(\omega_b(\mathcal{N}_{U}^2(1+2 N(\omega_U))(2\lambda-\omega_b)(\lambda+\omega_b-\sqrt{\omega_b(2\lambda+\omega_b)})+\mathcal{N}_{L}^2(1+2 N(\omega_L))(2 \lambda+\omega_b)(\lambda-\omega_b+\sqrt{\omega_b(-2 \lambda+\omega_b)})))/(16\omega_b(4 \lambda^2-\omega_b^2))$, and it means the covariance matrix of the subsystem is equal. The related coefficients can reduce to $w_{\pm}=\frac{1}{\mathcal{M}_{j}}(\frac{\omega_{a}+\sqrt{\omega_{a}(\omega_{a}\pm2 \lambda)}}{\lambda} \pm 1)$, $x_{\pm}=\frac{1}{\mathcal{M}_{j}}(1 \pm \frac{\omega_{a}+\sqrt{\omega(\omega_{a} \pm 2 \lambda)}}{\lambda})$, $y_{j}=\frac{\pm 1}{\mathcal{M}_{j}}$, and $z_{\pm}=\frac{1}{\mathcal{M}_{j}}$ with the normalized coefficients $\mathcal{M}_{j}=\frac{\sqrt{-2 \lambda^2+2(\lambda\pm \omega+\sqrt{\omega(\omega\pm 2 \lambda)})^2}}{\lambda}$. In this resonant case, one can find that $|w_j|=|x_j|$, and $|y_{j}|=|z_{j}|$. Hence, the one-way EPR steering does not exist in the system for resonant cases. To realize one-way EPR steering, some asymmetry needs to be considered. We here consider off-resonant case $\omega_a=2 \omega_b$, and the one-way EPR steering $\mathcal{G}^{a \rightarrow b}$ can be implemented, which is shown in Fig. \ref{fig8}.

\begin{widetext}
\section{The global master equation in the local representation} \label{Appendix D}

The global master equation as shown in Eq. (\ref{eq:common_ms}) in the local representation can read as 
\begin{align}
\begin{split}
\frac{\mathrm{d \rho}}{\mathrm{d} t}=&[\Gamma^{C}(-\omega_j) |w_j|^2+\Gamma^{C}(\omega_j) |y_j|^2] \mathcal{R}[a, a^{\dagger}] \rho+[\Gamma^{C}(-\omega_j) |x_j|^2+\Gamma^{C}(\omega_j) |z_j|^2]\mathcal{R}[b, b^{\dagger}] \rho+ \\&
[\Gamma^{C}(-\omega_j) |y_j|^2+\Gamma^{C}(\omega_j) |x_j|^2] \mathcal{R}[a^{\dagger}, a] \rho+[\Gamma^{C}(-\omega_j) |z_j|^2+\Gamma^{C}(\omega_j) |x_j|^2] \mathcal{R}[b^{\dagger}, b] \rho + \\&
[\Gamma^{C}(-\omega_j) w_j y_j^{*}+\Gamma^{C}(\omega_j) w_j y_j^{*}] \mathcal{R}[a, a] \rho+[\Gamma^{C}(-\omega_j) x_j z_j^{*}+\Gamma^{C}(\omega_j) x_j z_j^{*}] \mathcal{R}[b, b] \rho + \\&
[\Gamma^{C}(-\omega_j) w_j^{*} y_j+\Gamma^{C}(\omega_j) w_j^{*} y_j \mathcal{R}[a^{\dagger}, a^{\dagger}] \rho+[\Gamma^{C}(-\omega_j) z_j x_j^{*}+\Gamma^{C}(\omega_j) x_j z_j^{*}] \mathcal{R}[b^{\dagger}, b^{\dagger}] \rho + \\&
[\Gamma^{C}(-\omega_j) w_j x_j^{*}+\Gamma^{C}(\omega_j) z_j y_j^{*}] \mathcal{R}[a, b^{\dagger}] \rho+[\Gamma^{C}(-\omega_j) z_j y_j^{*}+\Gamma^{C}(\omega_j) w_j x_j^{*}]\mathcal{R}[b^{\dagger}, a] \rho +\\&
[\Gamma^{C}(-\omega_j) w_j z_j^{*}+\Gamma^{C}(\omega_j) w_j z_j^{*}] \mathcal{R}[a, b] \rho+[\Gamma^{C}(-\omega_j) x_j y_j^{*}+\Gamma^{C}(\omega_j) w_j z_j^{*}] \mathcal{R}[b, a] \rho + \\&
[\Gamma^{C}(-\omega_j) x_j w_j^{*}+\Gamma^{C}(\omega_j) y_j z_j^{*}] \mathcal{R}[b, a^{\dagger}] \rho+[\Gamma^{C}(-\omega_j) y_j z_j^{*}+\Gamma^{C}(\omega_j) x_j w_j^{*}] \mathcal{R}[a^{\dagger}, b] \rho + \\&
[\Gamma^{C}(-\omega_j) z_j w_j^{*}+\Gamma^{C}(\omega_j) y_j x_j^{*} \mathcal{R}[b^{\dagger}, a^{\dagger}] \rho+[\Gamma^{C}(-\omega_j) y_j x_j^{*}+\Gamma^{C}(\omega_j) z_j w_j^{*}] \mathcal{R}[a^{\dagger}, b^{\dagger}] \rho,  
\end{split}
\end{align}
\end{widetext}
where $\mathcal{R}[o, o^{\prime}] \rho=o \rho o^{\prime \dagger} -\frac{1}{2} \lbrace o^{\prime \dagger} o, \rho \rbrace$. 
For the Hopfield model without $A^2$ term, the coefficients of global generator in the local presentation are symmetry for modes $a$ and $b$ for resonant case as the condition $|w_j|=|x_j|$, and $|y_{j}|=|z_{j}|$. However, the Hopfield model does not satisfy this condition, one need to analyze the relation of the coefficients of global generator for modes $a$ and $b$ for resonant case. Without loss of generality, we compare the coupling coefficients of $\mathcal{R}[a, a^{\dagger}] \rho$ and $\mathcal{R}[b, b^{\dagger}] \rho$ and seek the condition of their equality, for example, $(\Gamma^{C}(-\omega_j) |w_j|^2+\Gamma^{C}(\omega_j) |y_j|^2)-(\Gamma^{C}(-\omega_j) |x_j|^2+\Gamma^{C}(\omega_j) |z_j|^2)=0$. Other cases are the same. After some algebras, the equation can become $(|x_{U}|^2-|x_{L}|^2)(\Gamma^{C}(-\omega_{U})-\Gamma^{C}(-\omega_{L}))+(|z_{U}|^2-|z_{L}|^2)(\Gamma^{C}(\omega_{U})-\Gamma^{C}(\omega_{L}))=0$. From Eqs. (\ref{Eq.coefficients-1}) and (\ref{Eq.coefficients-2}), one can find that $|x_{L}|^2=|w_{U}|^2$, $|z_{L}|^2=|y_{U}|^2$. This equation can be solved as $\frac{\sin^2 \theta(\frac{\omega_U}{\omega_b}+\frac{\omega_b}{\omega_U}+2)-\cos^2 \theta(\frac{\omega_U}{\omega_a}+\frac{\omega_a}{\omega_U}+2)}{4} (N (\omega_{U})-N(\omega_{L}))=0$, and  one can obtain the condition $\cos 2 \theta=0$ for resonant case. However, $\cos2 \theta=0$, it means that $D=0$. Hence the existence of diamagnetic term $D$ leads to the asymmetry of subsystem-reservoir coupling.   

\bibliography{correlation}
\end{document}